\documentclass[conference]{IEEEtran}

\usepackage{graphicx}
\usepackage{subfigure}
\usepackage{amsmath}
\usepackage{amssymb}
\usepackage{setspace}
\usepackage[compress]{cite}
\makeatletter
\renewcommand{\citepunct}{,\penalty\@m\hskip.13emplus.1emminus.1em}
\renewcommand{\citedash}{\hbox{--}\penalty\@m}
\makeatother

\begin{document}
\title{Impact of Channel Asymmetry on Base Station Cooperative Transmission with Limited Feedback}

\author
{Xueying Hou, and Chenyang Yang\\
\authorblockA{School of Electronics and Information Engineering, Beihang University, Beijing, China \\}

}

\maketitle

\begin{abstract}
Base station (BS) cooperative transmission, also known as
coordinated multi-point transmission (CoMP), is an effective way to
avoid inter-cell interference in universal frequency reuse cellular
systems. To gain the promised benefit, however, huge feedback
overhead is in demand to gather the channel information. In this
paper, we analyze the impact of channel asymmetry, which is inherent
in CoMP systems, on downlink BS cooperative transmission with
limited feedback. We analyze the per-user rate loss of a multi-user CoMP
system led by quantization. Per-cell quantization of multicell
channels is considered, which quantizes the local channel and cross
channel separately and is more feasible in practice. From both the
analytical and simulation results, we provide a whole picture on
various critical factors that lead to the performance loss.
Specifically, we show that the per user rate loss led by limited
feedback depends on the location of its paired users, except for
relying on its own signal to noise ratio and the quantization errors
as in single cell multi-user multiple antenna systems. This implies
that the quantization accuracy required for local and cross channel
of each user depends on the locations of its own as well as its
paired users.
\end{abstract}

\section{Introduction}

Base station (BS) cooperative transmission, which is also known as
coordinated multi-point transmission (CoMP), is an effective
strategy to mitigate the inter-cell interference (ICI) arisen from
universal frequency reuse cellular systems. As a promising transmit
strategy, coherent cooperative transmission can enhance the downlink
spectrum efficiency by using multiuser (MU) multiple-input
multiple-output (MIMO) precoding
\cite{Foschini-NetworkMIMO-06,Tolli08,Huang09}, when both data and
channel state information (CSI) can be sent to a central unit (CU)
via backhaul links.

For frequency division duplexing (FDD) MU-MIMO systems, the required
CSI at the transmitter can be obtained through uplink feedback using
limited number of bits \cite{Lov08}. The impact of limited feedback
on the performance of single cell MU-MIMO transmission under spatial
independent channels has been investigated in \cite{Jindal06} and
\cite{GoldSmith07}. It shows that the quantization error of limited
feedback leads to a throughput ceiling at high signal-to-noise ratio
(SNR) level \cite{Jindal06}. In order to achieve full multiplexing
gain, the feedback bits per mobile station (MS) should increase
linearly with SNR at a slope proportional to the number of transmit
antennas.

Despite that the considered CoMP system with the CU can be viewed as
a large MIMO system with a "super BS", there are distinct
differences between the channels of the single cell MIMO
transmission and CoMP. An inherent feature of CoMP channels is
\textit{asymmetry}, which means that the average channel energy from
different BSs to one MS are different. This leads to different
performance when the same transmit strategy is applied in the two
systems \cite{Choi08}. It also provides new opportunities for
developing new transmit strategies \cite{HY09}. For single user (SU)
MIMO CoMP, dynamic allocating the feedback bits among the channels
with different energies is shown to perform better than the equally
bits allocation through simulations \cite{Huawei10ICASSP}, and a
per-cell bit allocation algorithm based on user location  is
provided aiming to achieve a normalized SNR upper bound \cite{SY10}.
For coordinated beamforming in multicell multiuser system, the CSI
quantization and codebook design are studied in \cite{Heath09TSP}
and \cite{Heath10ICASSP}, where only CSI are shared among the BSs
thereby the addressed scheme is not the MU-MIMO CoMP. In
\cite{Heath09TSP}, an adaptive feedback bits assignment among
desired and interfering channels exploiting the difference of their
average energies is proposed, where the desired and the interfering
CSI are quantized using separate codebooks. A better CSI
quantization method is proposed in \cite{Heath10ICASSP}, which
quantizes the concatenation of the desired and interfering channel
vectors by choosing a codeword from a designed codebook.

In this paper, we analyze the impact of channel asymmetry on MU-MIMO
CoMP using multicell zero-forcing beamforming (ZFBF) with limited
feedback. We consider per-cell channel quantization since it is of
practical importance, which respectively quantizes the channel
direction information (CDI) of local channel, i.e., the channel
between the BS and the mobile station (MS) who are in the same cell,
and that of cross channels, i.e., the channels between the BS and
the MS who are in different cells. We analyze the rate loss of
MU-MIMO CoMP caused by limited feedback. We will show that the rate
loss not only depends on those factors dominating the rate loss of
single cell MU-MIMO, which include the CSI quantization error, the
number of transmit antennas and the receive SNR, but also relies on
the location of the paired MSs. The performance loss of a MS led by
limited feedback reduces significantly when its paired MSs locate
near their serving BSs. This is especially true when this MS does
not locate at the cell edge.

\textit{Notations:} Boldface upper case letters $\mathbf{X}$
represent matrices, boldface lower case letters $\mathbf{x}$
denote vectors and standard lower case letters $x$ denote scalars. $\mathbf{X}(:,i)$ represents the $i$th column of $\mathbf{X}$. The transpose and Hermitian conjugate transpose of $\mathbf{X}$ are denoted as $\mathbf{X}^T$ and $\mathbf{X}^H$, respectively. $\mathbb{E}\{x\}$ is the expectation of the
random variable $x$ . $|x|$ represents the norm of
the complex scalar $x$, and $\|\mathbf{x}\|$ represents the two-norm
of $\mathbf{x}$.

\section{System Model} \label{S:System_Model}
\begin{figure}
\center
\begin{minipage}[t]{0.4\textwidth}
\includegraphics[width=1\textwidth]{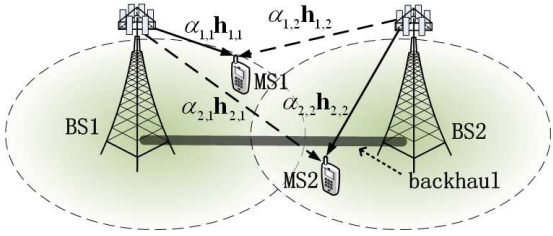}
\end{minipage}
\caption{\label{F:2CellSenario} An example of CoMP system with two
multiple antenna BSs cooperatively serving two single antenna MSs.
The solid lines denote the local channel while the dash lines
represent the cross channels.    }
\end{figure}

Consider a cellular system with $N$ cells, each contains one BS
equipped with $n_t$ antennas. To highlight the impact of channel
asymmetry on the performance of MU-MIMO CoMP, we assume that only
one single antenna MS will be selected from each cell for MU-MIMO
CoMP. When CSI from the $N$ BSs to the $N$ MSs are available in the
CU, MU-MIMO precoding is used for coherent cooperative transmission.
An example of CoMP system is shown in Fig.\ref{F:2CellSenario},
where either BS $1$ or BS $2$ can serve as the CU.

The global composite channel vector from all cooperative BSs to the
$k$th MS is represented by
\begin{equation}\label{E:Channel}
\mathbf{g}_k = [\alpha_{k,1}\mathbf{h}_{k,1}, \cdots, \alpha_{k,N} \mathbf{h}_{k,N}]
\end{equation}
where $\alpha_{k,b}$ and $\mathbf{h}_{k,b} \in \mathbb{C}^{1 \times
n_t }$ are respectively the large scale fading factor and the small
scale fading channel vector from the $b$th BS to the $k$th MS,
$\alpha_{k,b}$ includes path loss and shadowing, $\alpha_{k,b}^2$  is the
average energy of the composite channel
$\alpha_{k,b}\mathbf{h}_{k,b}$. For MS $k$ served by BS $k$,
$\alpha_{k,k}\mathbf{h}_{k,k}$ is its composite local channel
whereas $\alpha_{k,b}\mathbf{h}_{k,b}$ for $b \neq k$ are its
composite cross channels.

Let the data transmitting to all $N$ MSs be $\mathbf{d} = [d_1,
\cdots, d_N]^T$. Without loss of generality, we assume that
$\mathbb{E}\{\mathbf{d}\mathbf{d}^T\}= \mathbf{I}$. The received
signal at MS $k$ is
\begin{equation} \label{E:DL_RxSignal}
    y_k = \sqrt{P} \mathbf{g}_k \mathbf{v}_k d_k +  \sqrt{P} \sum_{j=1,j \neq k}^N  \mathbf{g}_k \mathbf{v}_j d_j+ n_k
\end{equation}
where $P$ is the transmit power for each MS, $\mathbf{v}_j$ is a
$Nn_t \times 1$ vector representing a unitary precoding vector from
all cooperative BSs for MS $j$, and $n_k$ is additive white Gaussian
noise (AWGN) with zero mean and variance $\sigma_n^2$.

Then we have the signal to noise plus interference ratio (SINR) at
MS $k$ as
\begin{equation}
\mathrm{SINR}_k = \frac{P|\mathbf{g}_k \mathbf{v}_k|^2}{\sigma_n^2 +
P \sum_{j=1,j \neq k}^N  |\mathbf{g}_k \mathbf{v}_j|^2 },
\end{equation}
and the average throughput of MS $k$ as
\begin{equation} \label{E:Throughput}
R_k = \mathbb{E}_{\mathbf{H},\mathbf{V}} \left\{ \log_2(1 + \mathrm{SINR}_k)\right\}.
\end{equation}

\subsection{Finite Rate Feedback Model}
We assume that MS $k$ has perfect knowledge of the global channel
vector $\mathbf{g}_k$. We consider \emph{per-cell quantization} to
quantize $\mathbf{g}_k$, i.e., MS $k$ separately quantizes the
channel vector $\alpha_{k,b}\mathbf{h}_{k,b}$ between MS $k$ and BS
$b$, $b=1,\cdots, N$. After MS $k$ feeds back these quantized
per-cell CSI, BS $k$ uses these feedback to reconstruct the global
channel of MS $k$ and then sends it to the CU.

Denote the instantaneous norm and CDI of the global channel between
MS $k$ and BS $b$ as $\rho_{k,b} = \alpha_{k,b}
\|\mathbf{h}_{k,b}\|$ and
$\mathbf{\bar{h}}_{k,b}=\mathbf{h}_{k,b}/\|\mathbf{h}_{k,b}\|$,
where $\rho_{k,b}$ includes both large scale fading factor and small
scale fading channel norm. To highlight the effect of CDI
quantization, we assume that the channel norm $\rho_{k,b}$,
$b=1,\cdots,N$, are directly fed back without quantization as in
\cite{Jindal06,GoldSmith07,Heath09TSP,Heath10ICASSP}. Denote the
codebook for quantizing the CSI between MS $k$ and BS $b$ as
$\mathcal{C}_{k,b}$, which is consist of unit norm row vectors
$\mathbf{c}_{i}$, $i=1,\cdots,2^{B_{k,b}}$. The quantized CDI
$\mathbf{\hat{h}}_{k,b}$ is set to be $\mathbf{c}_{i_{k,b}}$ with
the index $i_{k,b}$ chosen as follows
\begin{equation}
i_{k,b} = \mathrm{arg} \max_{1 \leq j \leq 2^{B_{k,b}}} |\mathbf{\bar{h}}_{k,b} \mathbf{c}_{j}^H|^2. \nonumber
\end{equation}
The quantization error is defined as $\sin^2 \theta_{k,b} = 1 -
|\mathbf{\bar{h}}_{k,b} \mathbf{\hat{h}}_{k,b}^H|^2$.

After MS $k$ quantizes the CDI for both local and cross channels, it
feeds back the index $\{i_{k,1},\cdots,i_{k,B}\}$ to its serving BS,
i.e., the BS $k$, which requires $\sum_{b=1}^N B_{k,b}$ bits. Based
on the index and the codebook, BS $k$ reconstructs the global
channel of MS $k$ as follows,
\begin{equation}\label{E:ConstructChannel}
\mathbf{\hat{g}}_k = [\rho_{k,1}\mathbf{\hat{h}}_{k,1}, \cdots, \rho_{k,N}\mathbf{\hat{h}}_{k,N}].
\end{equation}

When all $N$ BSs obtained the reconstructed global channel vectors,
they send the quantized channel vectors to the CU via a low latency
and error free backhaul.

\subsection{Multicell Zero-forcing Beamforming}
After the reconstructed global channel vectors from $N$ MSs to $N$
BSs are available at the CU, a multicell ZFBF is obtained as follows
\begin{equation}
\mathbf{V} = \mathbf{\hat{H}}^H \left( \mathbf{\hat{H}} \mathbf{\hat{H}}^H\right)^{-1}
\end{equation}
where $\mathbf{\hat{H}} = \left[ \mathbf{\hat{g}}_1^H,
\cdots,\mathbf{\hat{g}}_N^H \right]^H$. We consider per-user power
constraint as in \cite{Jindal06}. Then the beamforming vector of all
cooperative BSs for the $k$th MS is obtained by normalizing the
$k$th column of $\mathbf{V}$ as $\mathbf{v}_k =
\mathbf{V}(:,k)/\|\mathbf{V}(:,k)\|$.

\section{Rate loss analysis of CoMP transmission} \label{S:Analysis}
Similar to \cite{Jindal06,GoldSmith07}, we study the rate loss led
by the limited feedback which can be bounded by \cite{Jindal06},
\begin{equation} \label{E:deltaR}
\Delta R_k < \log_2\left[1+\frac{P}{\sigma_n^2} \mathbb{E}\left\{ \sum\nolimits_{j \neq k} |\mathbf{g}_k \mathbf{v}_j|^2\right\}\right]. \\
\end{equation}

The rate loss caused by limited feedback in single cell MU-MIMO
systems is well analyzed in \cite{Jindal06}. Under the assumption
that both the channel vector and its quantization are isotropic
distributed, which holds for spatial independent fading channels and
the channels are quantized by random vector quantization,
\cite{Jindal06} shows that the rate loss is dominated by the
quantization error, the number of transmit antennas and the receive
SNR. When the quantization error is fixed, the rate loss will
increase with SNR.

In CoMP systems, due to the inherent asymmetry feature of multicell
channels, the CDI distribution of the global channel is largely
dependent on the large scale fading factors. As a result, the CDI
distribution of the global channel is no longer isotropic. In the
following, we will show that the rate loss of a MS in MU-MIMO CoMP
systems not only relies on those factors which dominate the per MS
rate loss in single cell MU-MIMO systems, but also depends on the
location of its paired MSs. Contrary to the conclusion drawn in
single cell MU-MIMO systems \cite{Jindal06}, the rate loss in
MU-MIMO CoMP systems may even decrease with the increase of SNR.

To decouple the impact of limited feedback on scheduling and
precoding, we assume that MU MIMO scheduling is conducted before the
downlink CoMP based on the quantized global channels. We can use
many scheduling algorithms for pairing the multiple users, e.g., the
semi-orthogonal user selection (SUS) \cite{GoldSmithSUS06}, which
chooses MSs that have high channel qualities and are mutually
semi-orthogonal. Specifically, the correlation of the quantized
channels between the scheduled MS $k$ and MS $j$, which is defined
as $\frac{|\mathbf{\hat{g}}_k
\mathbf{\hat{g}}_j^H|}{\|\mathbf{\hat{g}}_k
\|\|\mathbf{\hat{g}}_j\|}$, should be less than a threshold.

The ZFBF consists of pseudo-inverse of the channel matrix, which is
rather involved for analysis. For mathematical tractability, we
assume that the selected MSs are mutually orthogonal in terms of
their quantized globe channel directions, i.e., we set the threshold
in SUS as $0$. In practice, this assumption holds when the number of
MSs is large enough. Although this assumption is too strong, we will
verify through simulation results in section IV that the following
analysis is applicable for the realistic cellular scenario without
the assumption.

Under this assumption, the precoder vector for the MS $j$ reduces to
$\mathbf{v}_j = \mathbf{\hat{g}}_j^H/\|\mathbf{\hat{g}}_j\|$. Then
the rate loss in (\ref{E:deltaR}) can be expressed as
\begin{equation} \label{E:deltaR1}
\Delta R_k < \log_2\left[1+\frac{P}{\sigma_n^2} \mathbb{E}\left\{ \sum\nolimits_{j \neq k}\frac{1}{\|\mathbf{\hat{g}}_j\|^2} |\mathbf{g}_k \mathbf{\hat{g}}_j^H|^2\right\}\right]. \\
\end{equation}

Substituting the expressions of the global channel vector
$\mathbf{g}_k$ and its per-cell quantization $\mathbf{\hat{g}}_k$
into (\ref{E:deltaR1}) and after some manipulations, we can obtain
the rate loss of MS $k$ as (see Appendix \ref{A:proof-1} for
details),
\begin{equation} \label{E:deltaR2}
\Delta R_k<\log_2[ 1+ \frac{n_t}{n_t-1} \sum\nolimits_{j \neq k} \underbrace{\sum\nolimits_{b=1}^N \beta_{j,b} \gamma_{k,b}^2 \sin^2 \theta_{k,b} }_{I_j}]
\end{equation}
where $\beta_{j,b} = \frac{\alpha_{j,b}^2}{\sum_{b=1}^N
\alpha_{j,b}^2 }$ only depends on the locations of its paired users,
$\gamma_{k,b}^2 = \frac{P\alpha_{k,b}^2}{\sigma_n^2}$ is the receive
SNR of the signal from BS $b$ to MS $k$, $\sin^2 \theta_{k,b}$ is
the per-cell quantization error of $\mathbf{\bar{h}}_{k,b}$.

The term $I_j$ in (\ref{E:deltaR2}) represents the inter-user
interference from MS $j$ to MS $k$. We can find that $I_j$ not only
depends on the channel quantization error, the number of transmit
antennas at each BS and the receive SNR of MS $k$, but also relies
on the value of $\beta_{j,b}$. This indicates that the rate loss of
MS $k$ is closely related to the positions of its paired MSs.

\begin{figure}
\begin{center}
\subfigure[MS2 locates at cell edge]{
\begin{minipage}[t]{0.22\textwidth}
\includegraphics[width=1.05\textwidth]{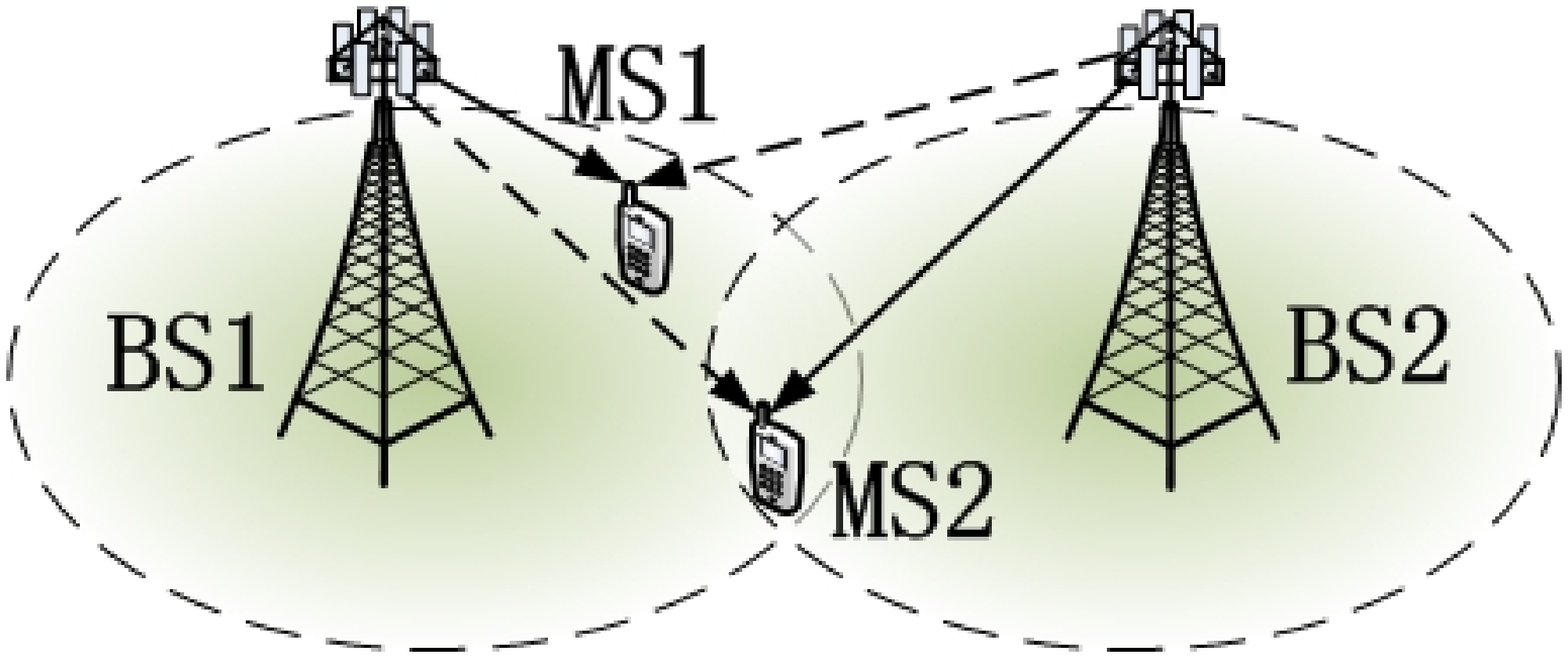}
\end{minipage}
}
\subfigure[MS2 locates at cell center]{
\begin{minipage}[t]{0.22\textwidth}
\includegraphics[width=1.05\textwidth]{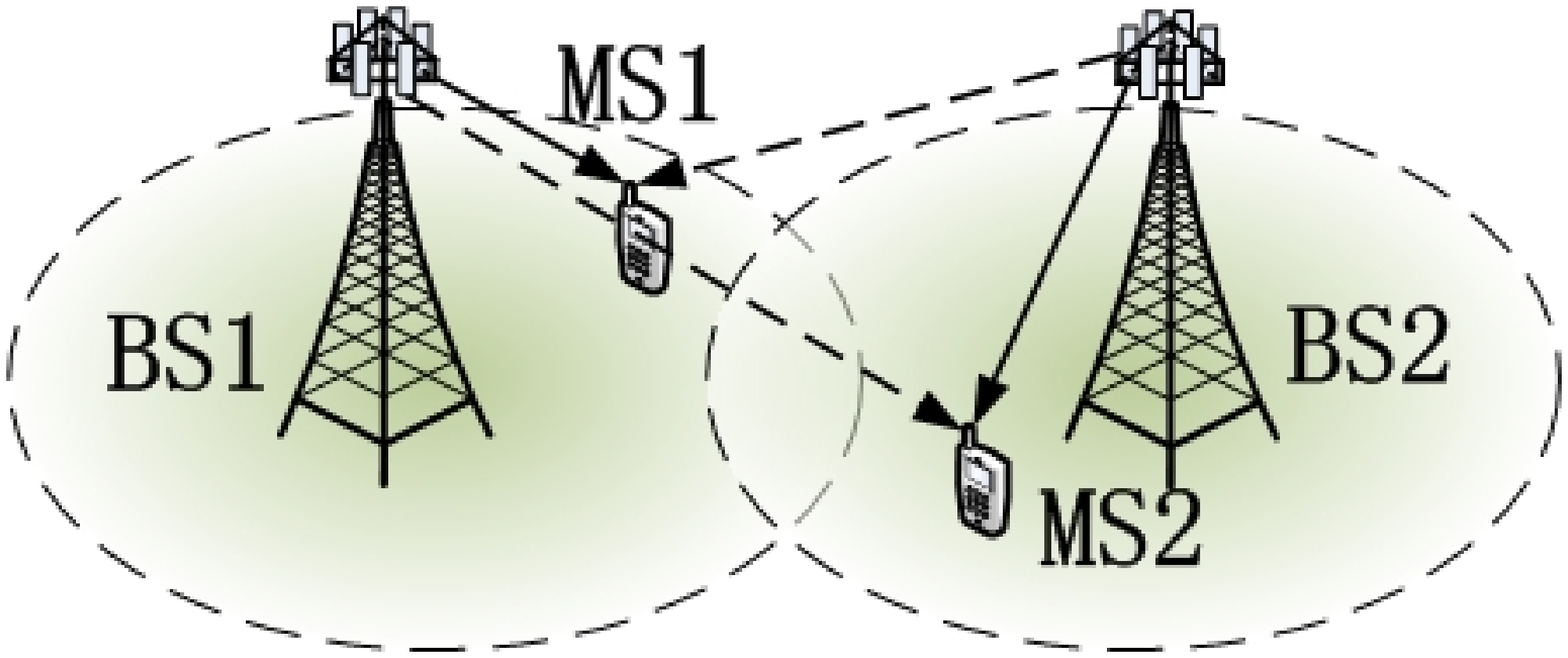}
\end{minipage}
}
\end{center}
\caption{\label{F:MS1Center} A CoMP system with two BSs cooperatively serving two MSs.}
\end{figure}

In order to provide a more clear picture on how the multicell MS
pairing will affect the rate loss, we consider a simple but
fundamental scenario where two BSs with multiple antennas
cooperatively serve two single-antenna MSs, as shown in
Fig.\ref{F:MS1Center}. Without loss of generality, we analyze the
rate loss of MS $1$, which is upper bounded by
\begin{align} \label{E:deltaR3}
\Delta R_1 & <\log_2\left[ 1+  \frac{n_t}{n_t-1} \left(\beta_{2,1} \gamma_{1,1}^2 \sin^2 \theta_{1,1} + \beta_{2,2} \gamma_{1,2}^2 \sin^2 \theta_{1,2}\right)\right] \nonumber\\
& \triangleq \Delta R_{1ub}
\end{align}
where $\beta_{2,1} = \frac{1}{1 + \alpha_{2,2}^2/\alpha_{2,1}^2 }$ and $\beta_{2,2} = \frac{1}{1 + \alpha_{2,1}^2/\alpha_{2,2}^2}$.

The value of $\gamma_{1,1}^2$ and $\gamma_{1,2}^2$ respectively
represent the local and cross link receive SNR of MS$1$, and $ \sin^2
\theta_{1,1}$ and $ \sin^2 \theta_{1,2}$ are the quantization
errors of the local CDI and the cross CDI. The values of
$\beta_{2,1}$ and $\beta_{2,2}$ depend on the average channel energy
ratio of the local channel and cross channel of MS$2$, i.e.
$\alpha_{2,2}^2/\alpha_{2,1}^2$. Here, MS2 is the paired MS for MS1.
To observe the impact of MS$2$ on the performance of MS$1$, we
analyze the following two cases when MS$2$ locates at cell edge and
at cell center.

\subsection{When MS$2$ Locates at Cell Edge}
As shown in Fig.\ref{F:MS1Center} (a), when MS $2$ locates at the
edge of cell $2$, we have $\alpha_{2,1}^2 = \alpha_{2,2}^2$. Then,
$\beta_{2,1}=\beta_{2,2}=\frac{1}{2}$. The upper bound of the rate
loss of MS $1$ becomes
\begin{equation}\label{E:deltaR4}
\Delta R_{1ub}  = \log_2\left[ 1+  \frac{n_t}{2(n_t-1)}
\left(\gamma_{1,1}^2\sin^2 \theta_{1,1} + \gamma_{1,2}^2 \sin^2
\theta_{1,2}\right)\right],
\end{equation}
which only depends on its receive SNRs and the per-cell CDI
quantization errors.

When MS$1$ also locates near the cell edge, i.e. , $\gamma_{1,1}^2
\approx \gamma_{1,2}^2$, this MU-MIMO CoMP scenario is similar to
the single cell MU-MIMO scenario. The rate loss depends on the
receive SNR and the per-cell quantization errors, where the
quantization errors of the local and cross channel contribute
equally to the rate loss. Moreover, due to the fact that the SNRs of
the local and cross links are low for the cell edge user MS1, the
rate loss caused by the limited feedback is insignificant since this
is a noise-limited scenario.

When MS$1$ moves from its cell edge to its cell center, the average
energy of its local channel will increase much faster than the
decrease of the average energy of its cross channel. Given the
feedback bits of the local channel and the cross channel, it is
shown from (\ref{E:deltaR4}) that the rate loss caused by the
limited feedback will increase dramatically. This means that the
throughput of the cell center MS is severely limited by the
quantization errors. Nonetheless, since $\gamma_{1,1}^2 \gg
\gamma_{1,2}^2$, the rate loss is mainly led by the quantization
error of the local channel, while that of the cross channel has
little contribution. This suggests that the rate loss can be reduced
by allocating more bits to the quantization of local channel than to
that of cross channel. Such a conclusion coincides with the
observations obtained for SU-MIMO CoMP \cite{Huawei10ICASSP} and
multi-user transmission using coordinated BF \cite{Heath09TSP}.

\subsection{When MS$2$ Locates at Cell Center}
As shown in Fig.\ref{F:MS1Center}(b), when MS $2$ locates at the
center of cell $2$, we have $\alpha_{2,2}^2 \gg \alpha_{2,1}^2$. As
a result, $\beta_{2,1}$ approaches $0$ while $\beta_{2,2}$ increases
towards $1$.

When MS$1$ locates near the cell edge, the value of $\gamma_{1,2}$
approximately equals to that of $\gamma_{1,1}$. At this time, the
contribution of the quantization error of the local channel is
little because $\beta_{2,1} \approx 0$. The rate loss of MS$1$ is
mainly caused by the quantization error of the cross channels. Since
the increase of $\beta_{2,2}$ is less than the decrease of
$\beta_{2,1}$, the rate loss of MS$1$ will be less when MS$2$
locates at cell center than that when MS$2$ locates at cell edge.

When MS$1$ locates near the cell center, we can observe from
(\ref{E:deltaR3}) that, when the receive SNR of its local channel is
high, the impact of the quantization error of the local channel is
less since $\beta_{2,1} \approx 0$. At the same time, the impact of
the quantization error of the cross channel will be less as well
since $\beta_{2,2} \leq 1$ and $\gamma_{1,2} \approx 0$. In general,
the rate loss caused by limited feedback will be neglected. This
implies that the feedback bits to both the local and the cross
channels can be significantly reduced if the MSs locate at the cell
center.

\section{Simulation results}
In this section, we verify our analysis through simulations. To be
consistent with our analysis, we consider two BSs cooperatively
serve two single antenna MSs. Each BS is equipped with $4$ antennas.
The cell radius $r$ is set to be $250$m. Assume that the receive SNR
of the cell edge MS, $\gamma_{edge}$, is $10$dB. Consider the path
loss factor $\epsilon$ as $3.76$, then the receive SNR of the signal
from a BS to a MS with MS-BS distance of $d$ can be calculated
according to $\gamma(d) = \gamma_{edge} +
\epsilon10\log_{10}(\frac{d}{r})$.

In the simulations, the small scale fading channels between BSs and
MSs are independent and identical distributed Rayleigh channels. The
codebooks for both the per-cell and the global channel quantization
are obtained by the generalized Lloyd algorithm
\cite{VectorQuantization}.

\subsection{Locations of MSs are Fixed}
Here we assume that two MSs locates on a straight line connecting
the two BSs. Then the distance from MS$1$ to BS$1$ can represent the
large scale fading factors of both local and cross channels of
MS$1$. When MS$1$ moves from its cell edge to its cell center, the
receive SNR of the local channel increases while that of the cross
channel decreases.

The throughput of MS$1$ is obtained according to
(\ref{E:Throughput}) by averaging over 1000 realizations of the
small scale fading channel. In order to observe if our analysis
still holds when the assumption of orthogonal MSs selection is not
satisfied, MS$2$ is always served simultaneously together with MS$1$
no matter whether its quantized global channel is orthogonal to that
of MS$1$ or not.

\begin{figure}
\center
\includegraphics[width=0.48\textwidth]{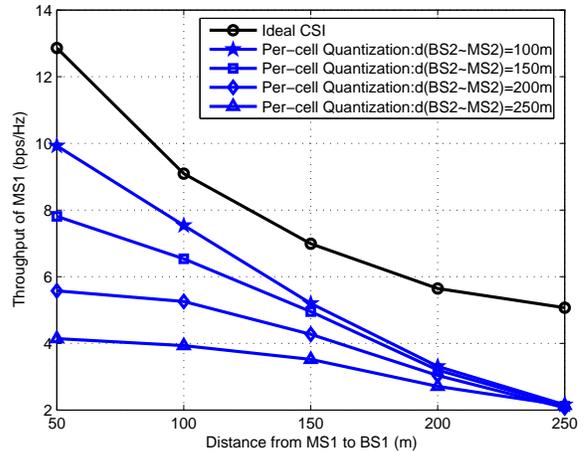}
\caption{\label{F:MS2Position} Throughput of MS1 versus the distance
from BS1 to MS1 when the distance from MS2 to BS2 are different.}
\end{figure}

\begin{figure}
\center
\includegraphics[width=0.48\textwidth]{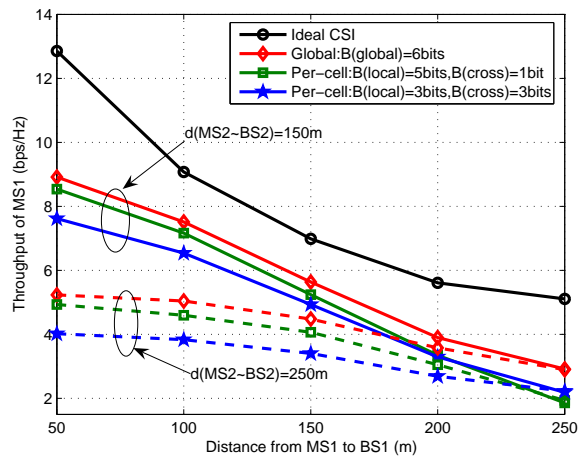}
\caption{\label{F:BitAllocation} Throughput of MS1 versus the
distance from BS1 to MS1 using different quantization methods.}
\end{figure}

Figure \ref{F:MS2Position} shows the average throughput of MS$1$
when the location of MS$2$ varies. Per-cell quantization is used and
the feedback bits of both local and cross channels are fixed to be
$3$ bits. To observe the rate loss, the performance with ideal CSI
is shown. Then the gap of the throughput using quantization from the
throughput using ideal CSI reflects the rate loss. It is shown that
when MS$2$ locates at cell edge, the average rate loss of MS$1$
increases when it moves from cell edge to cell center. Along with
the movement of MS$2$ towards BS$2$, the rate loss of MS$1$
decreases. When MS$2$ locates at the cell center, the performance
loss of MS$1$ even decreases when its local channel receive SNR
increases. This agrees well will our theoretical analysis.

Figure \ref{F:BitAllocation} shows the throughput of MS$1$ using
different quantization methods. Three quantization schemes are
considered, which are the global channel quantization, the per-cell
quantization with equal bits allocation and the per-cell
quantization with unequal bits allocation. It is shown that the
global quantization outperforms the per-cell quantization, and the
performance of per-cell quantization can be improved by unequal bits
allocation. These findings for the MU-MIMO CoMP coincide with those
for coordinate BF \cite{Heath10ICASSP}. However, we can observe that
the impact of the position of MS$2$ on the throughput of MS$1$ is
much larger than the quantization strategies.

\subsection{Locations of MSs are Randomly Distributed}
\begin{figure}
\center
\includegraphics[width=0.48\textwidth]{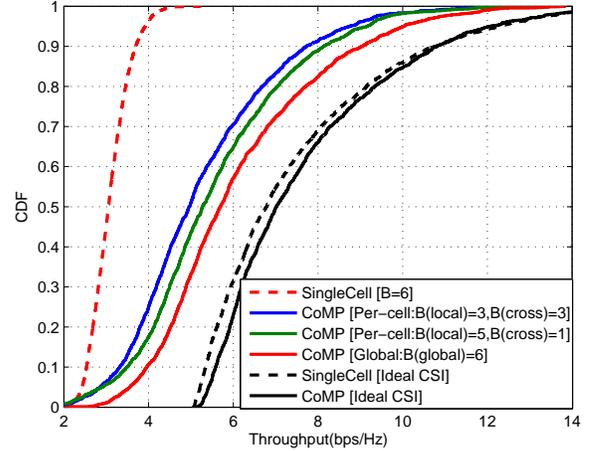}
\caption{\label{F:CDF_6Bits} CDF of per MS throughput when the two
MSs are randomly distributed. The impact of the limited feedback on
both single cell MU-MIMO and MU-MIMO CoMP systems are compared.}
\end{figure}

Now we simulate the case where the locations of two MSs are randomly
distributed in two cells, and the two MSs are served with MU-MIMO
CoMP by the two BSs. This corresponds to the scheme where a random
user selection method is applied. The cumulative distribution
function (CDF) of the throughput of MS$1$ is shown in
Fig.\ref{F:CDF_6Bits}, which is obtained from 1000 random drops. The
performances of the single cell MU-MIMO transmission with both ideal
CSI and quantized CSI are also shown for comparison. To ensure the
fairness of comparison with the CoMP scenario, two MSs are randomly
distributed in a single cell and the BS is equipped with $8$
antennas. Besides, we assume that the transmit power to the MSs are
the same for both single cell and CoMP scenarios. This means that
the BS transmit power of the single cell scenario is twice as much
as per-BS transmit power of the CoMP scenario.

We can observe that the performance of the single cell MU-MIMO
transmission degrades severely when $6$ bits are used to quantize
the $1 \times 8$ channel vectors. By contrast, the throughput gaps
between the schemes using limited feedback and ideal CSI are not
significant when only $6$ bits are used to feed back two $1 \times
4$ channel vectors in the CoMP system. This is due to the fact that,
for MU-MIMO CoMP, in a large probability the average channel energy
of the local channel for a randomly distributed MS is higher than
that of its cross channel. Therefore, the impact of quantization
error on the rate loss will be largely mitigated according to our
analysis in Section \ref{S:Analysis}.

\section{Conclusion}
In this paper, we investigated the impact of channel asymmetry on
CoMP MU-MIMO systems with limited feedback based on per-cell
quantization strategy. Our analysis showed that the rate loss of a
MS caused by limited feedback is dependent on the receive SNR, the
number of transmit antennas, the CSI quantization error, as well as
the location of its paired MSs in the cooperative cells. For a cell
center MS, the position of its paired MSs is the dominate factor of
its performance loss. When the paired MSs locates at cell edge, the
throughput of the cell center MS will be interference limited. When
the paired MSs also locates at cell center, the rate loss induced by
the quantization error is neglectable. On the other hand, for a cell
edge MS, its rate loss when its paired MSs locates at cell center
will also be less than that when the paired MSs locates at cell
edge.

\appendices
\section{Derivation of the upper bound of $\Delta R_k$} \label{A:proof-1}
Applying Jensen's inequality to (\ref{E:deltaR1}), we can obtain the
following upper bound of the rate loss of MS $k$,
\begin{align} \label{E:deltaR_A1}
\Delta R_k &< \log_2[1+\frac{P}{\sigma_n^2} \mathbb{E}\{ \sum\nolimits_{j \neq k}\frac{1}{\|\mathbf{\hat{g}}_j\|^2} |\mathbf{g}_k \mathbf{\hat{g}}_j^H|^2\}] \nonumber\\
&< \log_2[1+\frac{P}{\sigma_n^2} \sum_{j \neq k} \mathbb{E}\{ \frac{1}{\|\mathbf{\hat{g}}_j\|^2}\} \mathbb{E}\{ |\underbrace{\mathbf{g}_k \mathbf{\hat{g}}_j^H}_{Q_{j,b}}|^2\}].
\end{align}

Let us first find the upper bound of $\mathbb{E}\{
\frac{1}{\|\mathbf{\hat{g}}_j\|^2}\} $ in (\ref{E:deltaR_A1}) as
follows. Again by Jensen's inequality, we have
\begin{equation}\label{E:norm}
\mathbb{E}\{ \frac{1}{\|\mathbf{\hat{g}}_j\|^2}\}
 < \frac{1}{\sum_{b=1}^N  \mathbb{E} \{\rho_{j,b}^2\}}
 = \frac{1}{n_t\sum_{b=1}^N \alpha_{j,b}^2 }.
\end{equation}

Then before finding the upper bound of $\mathbb{E}\{ |Q_{j,b}|^2\}$
in (\ref{E:deltaR_A1}), we expand the expression of $Q_{j,b}$.
According to \cite{Jindal06}, the $\mathbf{h}_{k,b}$ can be
expressed as $\mathbf{h}_{k,b} = \cos \theta_{k,b}
\mathbf{\hat{h}}_{k,b} + \sin \theta_{k,b} \mathbf{s}_{k,b}$, where
$\mathbf{s}_{k,b}$ is isotropic distributed in the nullspace of
$\mathbf{\hat{h}}_{k,b}$. Then the expression of $Q_{j,b}$ can be
obtained as follows,
\begin{align}
Q_{j,b} = \sum\nolimits_{b=1}^N \varepsilon_{b}  \cos \theta_{k,b} \mathbf{\hat{h}}_{k,b}\mathbf{\hat{h}}_{j,b}^H +
 \sum\nolimits_{b=1}^N \varepsilon_{b}  \sin
 \theta_{k,b}\mathbf{s}_{k,b}\mathbf{\hat{h}}_{j,b}^H,\nonumber
\end{align}
where $\varepsilon_{b}= \rho_{k,b}\rho_{j,b}$.

Then we have
\begin{align} \label{E:A_E_Q}
\mathbb{E} \{|Q_{j,b}|^2\} =& \mathbb{E} \{ \sum\nolimits_{b=1}^N \varepsilon_b^2  \cos^2 \theta_{k,b} |\mathbf{\hat{h}}_{k,b}\mathbf{\hat{h}}_{j,b}^H|^2  \} \nonumber \\
&+\mathbb{E} \{\sum\nolimits_{b=1}^N \varepsilon_b^2  \sin^2 \theta_{k,b}|\mathbf{s}_{k,b}\mathbf{\hat{h}}_{j,b}^H|^2 \}  \nonumber \\
<&  \mathbb{E} \{ \sum\nolimits_{b=1}^N \varepsilon_b^2  |\mathbf{\hat{h}}_{k,b}\mathbf{\hat{h}}_{j,b}^H|^2  \} \nonumber \\
&+\mathbb{E} \{\sum_{b=1}^N \varepsilon_b^2  \sin^2
\theta_{k,b}|\mathbf{s}_{k,b}\mathbf{\hat{h}}_{j,b}^H|^2 \}.
\end{align}

Since we assume that the scheduled MSs are mutually orthogonal in
terms of quantized global CDI, we have
\[\mathbf{\hat{g}}_{k}\mathbf{\hat{g}}_{j}^H = \sum_{b=1}^N \varepsilon_b  \mathbf{\hat{h}}_{k,b}\mathbf{\hat{h}}_{j,b}^H=
0,\] where $\mathbf{\hat{g}}_{k}$ is the quantized global CDI of MS
$k$. Then by assuming that the channels from one BS to multiple MSs
are independent, we have
\begin{equation} \label{E:A_vertical}
\mathbb{E} \{|\mathbf{\hat{g}}_{k}\mathbf{\hat{g}}_{j}^H|^2\} =
\sum_{b=1}^N
\mathbb{E}\{\varepsilon_b^2|\mathbf{\hat{h}}_{k,b}\mathbf{\hat{h}}_{j,b}^H|^2\}
= 0.
\end{equation}

Substituting (\ref{E:A_vertical}) into (\ref{E:A_E_Q}) we can get
\begin{align}\label{E:A_E_Q1}
\mathbb{E} \{|Q_{j,b}|^2\}
& < \sum\nolimits_{b=1}^N \sin^2 \theta_{k,b} \mathbb{E}\{\varepsilon_b^2\}  \mathbb{E} \{|\mathbf{s}_{k,b}\mathbf{\hat{h}}_{j,b}^H|^2 \}\nonumber \\
& = \frac{1}{n_t-1}\sum\nolimits_{b=1}^N \alpha_{k,b}^2
\alpha_{j,b}^2  \sin^2 \theta_{k,b},
\end{align}
where $\mathbb{E} \{|\mathbf{s}_{k,b}\mathbf{\hat{h}}_{j,b}^H|^2 \} = \frac{1}{n_t-1}$ is obtained according to \cite{Jindal06}.

Substituting (\ref{E:norm}) and (\ref{E:A_E_Q1}) into (\ref{E:deltaR_A1}), we can obtain the upper bound of  the rate loss as in (\ref{E:deltaR2}).


\end{document}